\documentclass[prd,showpacs,amsmath,amssymb,superscriptaddress,nofootinbib]{revtex4}

\usepackage{graphicx}
\usepackage{dcolumn}
\usepackage{bm}

\newcommand{\Ref}[1]{(\ref{#1})}
\newcommand{\cqg}{Classical Quantum Gravity\ }
\newcommand{\grg}{Gen. Relativ. Gravit.\ }
\newcommand{\D}{\hat{D}}

\begin{document}

\title{Non-minimal Einstein-Yang-Mills-Higgs theory: \\
Associated, color and color-acoustic metrics for the Wu-Yang
monopole model}

\author{A. B. Balakin}
\email{Alexander.Balakin@ksu.ru} \affiliation{Department of
General Relativity and Gravitation, Kazan State University,
Kremlevskaya str. 18, Kazan 420008, Russia}
\author{H. Dehnen}
\email{Heinz.Dehnen@uni-konstanz.de} \affiliation{Universit\"at
Konstanz, Fachbereich Physik, Fach M 677, D-78457, Konstanz,
Germany}
\author{A. E. Zayats}%
\email{Alexei.Zayats@ksu.ru} \affiliation{Department of General
Relativity and Gravitation, Kazan State University, Kremlevskaya
str. 18, Kazan 420008,
Russia}%

\date{\today}

\begin{abstract}

We discuss a non-minimal Einstein-Yang-Mills-Higgs model with
uniaxial anisotropy in the group space associated with the Higgs
field. We apply this theory to the problem of propagation of color
and color-acoustic waves in the gravitational background related
to the non-minimal regular Wu-Yang monopole.

\end{abstract}


\maketitle

\section{Introduction}\label{Intro}

Non-minimal (induced by curvature) interactions of gravity field
with electromagnetic, gauge and scalar fields (see, e.g.,
\cite{FaraR,Hehl3} for review and basic references) are known to
introduce specific coupling constants, which can be associated
with radii of new (non-minimal) horizons, when one deals with
stationary spherically symmetric objects \cite{drummond,MHS,BL05}.
The causal structure of spacetime surrounding the charged objects
non-minimally coupled to gravity field is more sophisticated than
the standard (minimal) one. To illustrate this circumstance one
can mention recently obtained exact solutions of the
three-parameter non-minimal Einstein-Yang-Mills model for the
Wu-Yang monopole \cite{2BZ06} and Wu-Yang wormhole \cite{BSZ07}.
At the same time these instances show explicitly that an
appropriate choice of constants of non-minimal coupling, $q_1$,
$q_2$ and $q_3$, can result not only in appearance of new
(non-minimal) horizons, but in disappearance of singularities as
well. For example, when $q_1=-q$, $q_2=4q$, $q_3=-6q$ and $q$ is
positive, metric of the non-minimal Wu-Yang monopole (see
\cite{2BZ06}, Eq.(40)) is regular at the center $r=0$ and has no
horizons, when the mass of the object is less than the critical
one, $M_{({\rm crit})}$. In the presented paper we show that there
exists a simple extension of this regular solution to the case,
when the non-minimal Wu-Yang monopole possesses scalar Higgs
field, which is covariantly constant and parallel to the
Yang-Mills field strength in the group space. Considering the
gravitational, gauge and scalar fields of such monopole as a
non-minimal background, we start now to study the dynamics of {\it
test} particles, quasi-particles and waves in the monopole
environment. Despite the background gravity field is {\it
regular}, the interaction of particles and waves with curvature
(which can be reformulated in terms of effective non-minimal
forces) results in the appearance of singularities of a dynamic
type, analogous to the ones well-known in the Analogue Gravity
theory \cite{Visser1,Novello,VolovikBook}. For instance,
particles, quasi-particles and waves can fall within trapped
regions in the environment of the non-minimal Wu-Yang monopole.
The main instrument of analysis of such a problem is the theory of
effective metrics (see, e.g., the review \cite{Visser1}). It is
well known, that optical metrics are alternative tools to
investigate the propagation of light in the framework of
electrodynamics of moving media (see \cite{Gordon,PMQ} and, e.g.,
\cite{Perlick,HehlObukhov} for a review), the acoustic metrics
give an alternative description of propagation of scalar waves in
the framework of hydrodynamics \cite{Visser1,Novello,VolovikBook}.
When we deal with Yang-Mills-Higgs theory the {\it color} metrics
appear instead of optical ones, and {\it color-acoustic} metrics
can be introduced for the description of the Higgs scalar waves.
The study and physical interpretation of singularities, appearing
in color and color-acoustic metrics in the background of the
regular non-minimal Wu-Yang monopole solution, is the main goal of
the presented paper.

The paper is organized as follows. Section \ref{model} contains
description of the seven-parameter non-minimal
Einstein-Yang-Mills-Higgs (EYMH) model: Subsection \ref{lagr}
includes basic definitions, in Subsection \ref{higgs} we introduce
uniaxial structure in the group space and in Subsection
\ref{master} we obtain non-minimal master equations for the
Yang-Mills, Higgs and gravity fields. Section \ref{ex} is devoted
to the description of the exact solution of presented model,
supplemented by the Wu-Yang ansatz \cite{WuYang}. In Section
\ref{repr} we study associated metrics (\ref{assoc}), color
metrics (\ref{color}) and color-acoustic metrics (\ref{acoustic})
in the background of regular solution for the non-minimal Wu-Yang
monopole. In Discussion we consider in more details the admissible
and trapped regions for longitudinal and transversal color and
color-acoustic waves.

\section{Seven-parameter non-minimal EYMH model}\label{model}

\subsection{Lagrangian}\label{lagr}

Consider a modified action functional\footnote{Hereafter we use the
units $c=G=\hbar=1$.}
$$
S_{({\rm NMEYMH})} = \int d^4 x \sqrt{-g}\ \left\{
\frac{R}{8\pi}+\frac{1}{2}F^{(a)}_{ik} F^{ik}_{(a)}
-{\D}_m\Phi^{(a)}{\D}^m\Phi_{(a)} + \frac{1}{2} \lambda \left(
\Phi^2- v^2 \right)^2 + {}\right.
$$
\begin{equation}\label{1act}
\left. {}+ \frac{1}{2} {\cal R}^{ikmn} X_{(a)(b)} F^{(a)}_{ik}
F^{(b)}_{mn} - \Re^{mn} Y_{(a)(b)} (\D_m \Phi)^{(a)} (\D_n
\Phi^{(b)}) \right\}\,,
\end{equation}
where the susceptibility tensors ${\cal R}^{ikmn}$ and
$\Re^{\,mn}$ are defined as
\begin{equation}
{\cal R}^{ikmn} \equiv \frac{q_1}{2}R\,(g^{im}g^{kn}-g^{in}g^{km})
+ \frac{q_2}{2}(R^{im}g^{kn} - R^{in}g^{km} + R^{kn}g^{im}
-R^{km}g^{in}) + q_3 R^{ikmn}\,, \label{sus}
\end{equation}
\begin{equation}\label{Reqq}
\Re^{\,mn}\equiv {q_4}Rg^{mn}+q_5 R^{mn}\,.
\end{equation}
Here $g = {\rm det}(g_{ik})$ is the determinant of a metric tensor
$g_{ik}$, $R$ is the Ricci scalar, tensor $F^{(a)}_{ik}$ is the
strength of gauge field, the symbol $\Phi^{(a)}$ denotes the
multiplet of the Higgs scalar fields, $\frac{1}{2} \lambda \left(
\Phi^2-v^2 \right)^2$ is a potential of the Higgs field, and
$\Phi^2\equiv \Phi^{(a)}\Phi_{(a)}$. Latin indices without
parentheses run from 0 to 3, $(a)$ and $(b)$ are the group indices.
Following \cite{Rubakov}, Section 4.3, we consider the Yang-Mills
field ${\bf F}_{mn}$ and the Higgs field ${\bf \Phi}$ taking values
in the Lie algebra of the gauge group $SU(n)$:
\begin{equation}
{\bf F}_{mn} = - i {\cal G} {\bf t}_{(a)} F^{(a)}_{mn} \,, \quad
{\bf A}_m = - i {\cal G} {\bf t}_{(a)} A^{(a)}_m \,, \quad {\bf
\Phi} = {\bf t}_{(a)} \Phi^{(a)} \,. \label{represent}
\end{equation}
Here ${\bf t}_{(a)}$ are Hermitian traceless generators of $SU(n)$
group, $F^{(a)}_{mn}$, $A^{(a)}_i$ and $\Phi^{(a)}$ are real
fields ($A^{(a)}_i$ represents the Yang-Mills field potential),
${\cal G}$ is a constant of gauge interaction, and the group index
$(a)$ runs from $1$ to $n^2-1$. The scalar product of the
generators ${\bf t}_{(a)}$ and ${\bf t}_{(b)}$ is defined via the
trace
\begin{equation}
\left( {\bf t}_{(a)} , {\bf t}_{(b)} \right) \equiv 2 {\rm Tr} \
{\bf t}_{(a)} {\bf t}_{(b)}  \equiv G_{(a)(b)}  \,,
\label{scalarproduct}
\end{equation}
the symmetric tensor $G_{(a)(b)}$ plays a role of a metric in the
group space. The representation \Ref{represent},
\Ref{scalarproduct} allows to consider the multiplets of the real
fields $\{F^{(a)}_{mn}\}$, $ \{A^{(a)}_i\}$ and $\{\Phi^{(a)} \}$
as components of the corresponding vectors in the $n^2-1$
dimensional group space. The operations with the group indices
$(a)$ are the following: the repeating indices denote the
convolution, and the rule $\Phi_{(a)} = G_{(a)(b)} \Phi^{(b)}$ for
the indices lowering takes place.
The Yang-Mills fields
$F^{(a)}_{mn}$ are connected with the potentials of the gauge field
$A^{(a)}_i$ by the well-known formula \cite{Rubakov}
\begin{equation}
F^{(a)}_{mn} = \nabla_m A^{(a)}_n - \nabla_n A^{(a)}_m + {\cal G}
f^{(a)}_{\ (b)(c)} A^{(b)}_m A^{(c)}_n \,. \label{Fmn}
\end{equation}
Here $\nabla _m$ is a  covariant space-time derivative, the
symbols $f^{(a)}_{\ (b)(c)}$ denote the real structure constants
of the gauge group $SU(n)$. The gauge invariant derivative $\D_m
\Phi^{(a)}$ is defined according to the formula \cite{Rubakov}
\begin{equation}
\D_m \Phi^{(a)} \equiv \nabla_m \Phi^{(a)} + {\cal G}
f^{(a)}_{\cdot (b)(c)} A^{(b)}_m \Phi^{(c)} \,. \label{DPhi}
\end{equation}
The definition of the commutator is based on the relation:
\begin{equation}
\left[ {\bf t}_{(a)} , {\bf t}_{(b)} \right] = i  f^{(c)}_{\
(a)(b)} {\bf t}_{(c)} \,, \label{fabc}
\end{equation}
providing the formula
\begin{equation}
f_{(c)(a)(b)} \equiv G_{(c)(d)} f^{(d)}_{\ (a)(b)} = - 2 i \ {\rm
Tr} \ \left[ {\bf t}_{(a)} , {\bf t}_{(b)} \right] {\bf t}_{(c)}
\,. \label{fabc1}
\end{equation}
The structure constants $f_{(a)(b)(c)}$ are supposed to be
antisymmetric under exchange of any two indices. The quantities
$X_{(a)(b)}$ and $Y_{(a)(b)}$ are symmetric with respect to the
group indices and depend on the Higgs field $\Phi^{(c)}$; they are
defined below.

\subsection{Higgs field and uniaxial structures in the group
space}\label{higgs}

The group space can be considered as isotropic, when it is
equipped by the metric $G_{(a)(b)}$ only, and as anisotropic one
in all other cases. The presence of the Higgs field multiplet
$\Phi^{(a)}$ with positive norm $G_{(a)(b)}\Phi^{(a)}\Phi^{(b)}
\equiv  \Phi^2 > 0$ allows us to equip the group space by the
vector $q^{(a)} = \Phi^{(a)}/ \Phi$ with the unit norm $G_{(a)(b)}
q^{(a)} q^{(b)}=1$ and by the projector $P_{(a)(b)}$. The latter
is defined as
\begin{equation}
P_{(a)(b)} \equiv G_{(a)(b)} - \frac{\Phi_{(a)}
\Phi_{(b)}}{\Phi^2} = G_{(a)(b)} - q_{(a)}q_{(b)}\,,
\label{7para6}
\end{equation}
and possesses the evident projector properties:
\begin{equation}
P_{(a)(b)} =  P_{(b)(a)} \,, \quad P_{(a)(b)} \Phi^{(b)} = 0 \,,
\quad P_{(a)(b)} P^{(a)(c)} = P^{(c)}_{(b)} \,, \quad
P^{(a)}_{(a)} = n^2 - 2 \,. \label{7para7}
\end{equation}
Using these quantities, every tensor ${\cal A}^{i_1 i_2
...i_s}_{(a)(b)}$, symmetric with respect to the transposition of
the group indices $(a)$ and $(b)$, can be decomposed into the sum
\begin{equation}
{\cal A}^{i_1 i_2 ...i_s}_{(a)(b)} = {\cal A}^{i_1 i_2
...i_s}_{({\rm long})} q_{(a)}q_{(b)} +  {\cal B}^{i_1 i_2
...i_s}_{(a)} q_{(b)} + {\cal B}^{i_1 i_2 ...i_s}_{(b)} q_{(a)} +
{\cal D}^{i_1 i_2 ...i_s}_{(a)(b)}\,, \label{ekh}
\end{equation}
where
\begin{equation}
{\cal A}^{i_1 i_2 ...i_s}_{({\rm long})} \equiv  {\cal A}^{i_1 i_2
...i_s}_{(c)(d)} q^{(c)} q^{(d)} \,, \quad  {\cal D}^{i_1 i_2
...i_s}_{(a)(b)}  \equiv {\cal A}^{i_1 i_2 ...i_s}_{(c)(d)}
P^{(c)}_{(a)} P^{(d)}_{(b)} \,, \label{ekh1}
\end{equation}
\begin{equation}
{\cal B}^{i_1 i_2 ...i_s}_{(a)} \equiv {\cal A}^{i_1 i_2
...i_s}_{(c)(d)} P^{(c)}_{(a)} q^{(d)}  = {\cal A}^{i_1 i_2
...i_s}_{(c)(d)} q^{(c)} P^{(d)}_{(a)}   \,, \label{ekh2}
\end{equation}
\begin{equation}
{\cal B}^{i_1 i_2 ...i_s}_{(a)} q^{(a)} = 0 \,, \quad {\cal
D}^{i_1 i_2 ...i_s}_{(a)(b)} q^{(b)} = 0= {\cal D}^{i_1 i_2
...i_s}_{(a)(b)} q^{(a)}\,. \label{ekh3}
\end{equation}
The simplest type of anisotropy in the group space is the uniaxial
one. This case assumes that the only direction pointed by
$q^{(a)}$ is the selected one, while other orthogonal directions
are equivalent, i.e.,
\begin{equation}\label{eigen}
    {\cal A}^{i_1 i_2
    ...i_s}_{(a)(b)}q^{(b)}= {\cal S}_1^{i_1 i_2
    ...i_s} q_{(a)}\,,\quad {\cal A}^{i_1 i_2
    ...i_s}_{(a)(b)}P^{(b)(c)} ={\cal S}_2^{i_1 i_2
    ...i_s} P_{(a)}^{(c)}\,,
\end{equation}
where ${\cal S}_1^{i_1 i_2...i_s}$ and ${\cal S}_2^{i_1
i_2...i_s}$ are some non-equal tensors without group indices. The
conditions \Ref{eigen} mean that
\begin{equation}
{\cal B}^{i_1 i_2 ...i_s}_{(a)} = 0 \,, \quad {\cal D}^{i_1 i_2
...i_s}_{(a)(b)} = {\cal D}^{i_1 i_2 ...i_s}_{({\rm trans})}
P_{(a)(b)}\,, \quad {\cal D}^{i_1 i_2 ...i_s}_{({\rm trans})} =
\frac{1}{(n^2-2)} {\cal A}^{i_1 i_2 ...i_s}_{(a)(b)} P^{(a)(b)}
\,. \label{ekh4}
\end{equation}
Keeping in mind this features, we assume the tensors $X_{(a)(b)}$
and $Y_{(a)(b)}$ in the Lagrangian \Ref{1act} to have the simplest
form
\begin{equation}
X_{(a)(b)} \equiv G_{(a)(b)} + (Q_1-1)
\frac{\Phi_{(a)} \Phi_{(b)}}{\Phi^2} \,, \quad Y_{(a)(b)} \equiv
G_{(a)(b)} + (Q_2-1) \frac{\Phi_{(a)} \Phi_{(b)}}{\Phi^2} \,,
\label{XYZ}
\end{equation}
with two new coupling constants $Q_1$ and $Q_2$. In principle, we
can treat they as quantities, which depend on the vacuum value $v$
of the scalar field. When $Q_1=Q_2=1$, the tensors $X_{(a)(b)}$
and $Y_{(a)(b)}$ coincide with the metric $G_{(a)(b)}$ in the
group space. The introduction of $Q_1 \neq 1$ and/or $Q_2 \neq 1$
modifies the master equations for the gauge, scalar and
gravitational fields. When $Q_1=0$, $X_{(a)(b)}=P_{(a)(b)}$, and
when $Q_2=0$, $Y_{(a)(b)}=P_{(a)(b)}$. In general case
$X_{(a)(b)}$ (and $Y_{(a)(b)}$, correspondingly) consists of two
parts, parallel to the direction pointed by the scalar field, and
orthogonal to it, the longitudinal and transversal parts being,
respectively
\begin{equation}
X_{({\rm long})} = X_{(a)(b)} q^{(a)}q^{(b)} = Q_1 \,, \quad
X_{({\rm trans})} = \frac{1}{(n^2-2)} X_{(a)(b)} P^{(a)(b)} = 1
\,. \label{XY01}
\end{equation}
In addition, every vector in the group space, say, $A_k^{(a)}$,
can be decomposed using longitudinal and transversal components as
follows
\begin{equation}
A_k^{(a)} = q^{(a)} A_{k ({\rm long})} + A_{k ({\rm trans})}^{(a)}
\,, \label{XY011}
\end{equation}
where
\begin{equation}
A_{k ({\rm long})} \equiv A_k^{(a)}q_{(a)} \,, \quad
A^{(a)}_{k({\rm trans})} \equiv  P^{(a)}_{(b)} A_k^{(b)} \,.
\label{XY017}
\end{equation}

\subsection{Non-minimal master equations}\label{master}

\subsubsection{Non-minimal extension of the Yang-Mills field equations}
The variation of the action functional over the Yang-Mills
potential $A^{(a)}_i$ yields
\begin{equation}
\D_k {\cal H}^{ik}_{(a)}  =   {\cal G} (\D_k
\Phi^{(d)})f^{(b)}_{\cdot (a)(h)} \Phi^{(h)}\left[ G_{(b)(d)}
g^{ik} + Y_{(b)(d)} \Re^{ik}  \right] \,, \label{Heqs}
\end{equation}
where the tensor ${\cal H}^{ik}_{(a)}$ is
\begin{equation}
{\cal H}^{ik}_{(a)} = C^{ikmn}_{(a)(b)} F_{mn}^{(b)} = \left[
\frac{1}{2}( g^{im} g^{kn} - g^{in} g^{km}) G_{(a)(b)} + {\cal
R}^{ikmn} X_{(a)(b)} \right] F^{(b)}_{mn} \,. \label{HikR}
\end{equation}
Thus, the linear response tensor $C^{ikmn}_{(a)(b)}$ is now of the
form
\begin{equation}
C^{ikmn}_{(a)(b)} \equiv \left[ \frac{1}{2}( g^{im} g^{kn} -
g^{in} g^{km}) + {\cal R}^{ikmn} \right] G_{(a)(b)} + (Q_1-1)
\frac{\Phi_{(a)} \Phi_{(b)}}{\Phi^2} {\cal R}^{ikmn}   \,,
\label{HikR2}
\end{equation}
the longitudinal and transversal parts being, respectively
\begin{equation}
C^{ikmn}_{({\rm long})}  = \left[ \frac{1}{2}( g^{im} g^{kn} -
g^{in} g^{km}) + Q_1 {\cal R}^{ikmn} \right] \,, \label{lo1}
\end{equation}
\begin{equation}
C^{ikmn}_{({\rm trans})} =  \left[ \frac{1}{2}( g^{im} g^{kn} -
g^{in} g^{km}) + {\cal R}^{ikmn} \right] \,. \label{tr1}
\end{equation}
Mention that $C^{ikmn}_{({\rm trans})}$ can be obtained from
$C^{ikmn}_{({\rm long})}$ by the formal replacement $Q_1 \to 1$,
and let us use this features below for the simplifications of the
formulas. The color permittivity, impermeability and cross-effect
tensors are also anisotropic in the group space
\begin{equation}
\varepsilon^{im}_{(a)(b)} \equiv 2C^{ikmn}_{(a)(b)}U_kU_n =
\left[\Delta^{im}+ 2 {\cal R}^{ikmn} U_k U_n  \right]G_{(a)(b)} +
2 (Q_1-1) {\cal R}^{ikmn} U_k U_n \frac{\Phi_{(a)}
\Phi_{(b)}}{\Phi^2} \,, \label{Re}
\end{equation}
\begin{equation}
(\mu^{-1})^{im}_{(a)(b)} \equiv -2\
{}^*C^{*\,ikmn}_{(a)(b)}U_kU_n= \left[\Delta^{im} - 2 \
{}^{*}{\cal R}^{*\,ikmn} U_k U_n \right] G_{(a)(b)} - 2 (Q_1-1) \
{}^{*}{\cal R}^{*\,ikmn} U_k U_n \frac{\Phi_{(a)}
\Phi_{(b)}}{\Phi^2} \,, \label{Rmu}
\end{equation}
\begin{equation}
\nu^{im}_{(a)(b)} \equiv 2\ {}^*C^{ikmn}U_kU_n = {}^{*}{\cal
R}^{ikmn} U_k U_n \left[G_{(a)(b)} + (Q_1-1) \frac{\Phi_{(a)}
\Phi_{(b)}}{\Phi^2} \right] \,. \label{Rnu}
\end{equation}
Here the asterisk denotes the dualization procedure:
\begin{equation}\label{dualization}
    {}^*W^{ikmn}=\frac{1}{2}\epsilon^{ikpq}{W_{pq}}^{mn}\,,\qquad
    {}^*W^{*\,ikmn}=\frac{1}{2}\epsilon^{pqmn}\,{{}^*W^{ik}}_{pq}\,,
\end{equation}
$U^i$ and $\Delta^{im}\equiv g^{im}-U^iU^m$ are a unit timelike
vector ($U_iU^i=1$) and the projection tensor, respectively.
 The corresponding longitudinal and transversal components
of $\varepsilon^{im}_{(a)(b)}$, $(\mu^{-1})^{im}_{(a)(b)}$ and
$\nu^{im}_{(a)(b)}$ can be easily written in analogy with
\Ref{lo1} and \Ref{tr1}.

\subsubsection{Non-minimal extension of the Higgs field equations}

The variation of the action $S_{({\rm NMEYMH})}$ over the Higgs
scalar field $\Phi^{(a)}$ yields
$$
\D_m \left\{ \left[ g^{mn} G_{(a)(b)} + \Re^{mn} Y_{(a)(b)}
\right] \D_n \Phi^{(b)} \right\} = - \lambda \Phi_{(a)} \left(
\Phi^2 - v^2 \right) -
$$
\begin{equation}
{} - \frac{(Q_1-1)}{2\Phi^2} \ {\cal
R}^{ikmn}F^{(c)}_{ik}F^{(b)}_{mn} \Phi_{(b)} P_{(a)(c)} +
\frac{(Q_2-1)}{\Phi^2} \ \Re^{mn} P_{(a)(c)} \Phi_{(b)} (\D_m
\Phi^{(c)}) (\D_n \Phi^{(b)}) \,. \label{higgs1}
\end{equation}
The tensor ${\cal C}^{ik}_{(a)(b)}$, introduced as
\begin{equation}
{\cal C}^{ik}_{(a)(b)} = \left[g^{ik} + \Re^{ik} \right]G_{(a)(b)} +
(Q_2-1)\Re^{ik} \frac{\Phi_{(a)} \Phi_{(b)}}{\Phi^2} \,,
\label{higgs111}
\end{equation}
can also be decomposed into longitudinal and transversal
components
\begin{equation}
{\cal C}^{ik}_{(a)(b)} = {\cal C}^{ik}_{({\rm long})} q_{(a)}
q_{(b)} + P_{(a)(b)}{\cal C}^{ik}_{({\rm trans})} \,, \label{hi1}
\end{equation}
where
\begin{equation}
{\cal C}^{ik}_{({\rm long})} \equiv {\cal C}^{ik}_{(a)(b)}q^{(a)}
q^{(b)} \,, \quad {\cal C}^{ik}_{({\rm trans})} \equiv
\frac{1}{n^2-2}{\cal C}^{ik}_{(a)(b)} P^{(a)(b)} \,.  \label{hi2}
\end{equation}
The definitions
\begin{equation}
\tilde{g}^{ik}_{({\rm long})} \equiv {\cal C}^{ik}_{({\rm long})}
= g^{ik} + Q_2 \Re^{ik} \,, \label{hi3}
\end{equation}
\begin{equation}
\tilde{g}^{ik}_{({\rm trans})} \equiv {\cal C}^{ik}_{({\rm
trans})} = g^{ik} + \Re^{ik} \,, \label{hi4}
\end{equation}
introduce two color-acoustic metrics for the colored scalar
particles. Again, $\tilde{g}^{ik}_{({\rm trans})}$ can be obtained
from $\tilde{g}^{ik}_{({\rm long})}$ by the formal replacement
$Q_2 \to 1$.

\subsubsection{Master equations for the gravitational field}

The equations for the gravity field related to the action
functional $S_{({\rm NMEYMH})}$ are of the form
\begin{equation}
R_{ik} - \frac{1}{2} R \ g_{ik} = 8\pi T^{({\rm eff})}_{ik} \,,
\label{Ein}
\end{equation}
where
\begin{equation} T^{({\rm eff})}_{ik} =  T^{(YM)}_{ik} +
T^{(H)}_{ik} + T^{(NM)}_{ik} \,. \label{Tdecomp}
\end{equation}
The first term $T^{(YM)}_{ik}$:
\begin{equation}
T^{(YM)}_{ik} \equiv \frac{1}{4} g_{ik} F^{(a)}_{mn}F^{mn}_{(a)} -
F^{(a)}_{in}F_{k (a)}^{\cdot n} \,, \label{TYM}
\end{equation}
is a stress-energy tensor of pure Yang-Mills field, the second
term:
\begin{equation}
T^{(H)}_{ik} \equiv -  \frac{1}{2} g_{ik} \D_m \Phi^{(a)} \D^m
\Phi_{(a)} + \D_i \Phi^{(a)} \D_k \Phi_{(a)} + \frac{1}{4} g_{ik}
\lambda \left(\Phi^2 - v^2 \right)^2 \label{TH}
\end{equation}
relates to the standard stress-energy of the Higgs field. The
non-minimal contributions enter the last tensor $T^{(NM)}_{ik}$,
which may be represented as a sum of 5 items:
\begin{equation}
T^{(NM)}_{ik} \equiv  q_1 T^{(I)}_{ik} + q_2 T^{(II)}_{ik} + q_3
T^{(III)}_{ik} + q_4 T^{(IV)}_{ik} + q_5 T^{(V)}_{ik} \,.
\label{Tdecomp1}
\end{equation}
The definitions of these five tensors relate to the corresponding
coupling constant $q_1$, $q_2,\ \dots,\ q_5$. The tensors
$T^{(I)}_{ik}$, $T^{(II)}_{ik}$, \dots, $T^{(V)}_{ik}$ are
$$
T^{(I)}_{ik} = R X_{(a)(b)}
\left[\frac{1}{4}g_{ik}F^{(a)}_{mn}F^{mn (b)} -
F^{(a)}_{im}F_{k}^{\ m (b)} \right] -  \frac{1}{2} R_{ik}
X_{(a)(b)} F^{(a)}_{mn}F^{mn (b)} +
$$
\begin{equation}
+ \frac{1}{2} \left[ \D_{i} \D_{k} - g_{ik} \D^l \D_l \right]
\left[X_{(a)(b)}F^{(a)}_{mn}F^{mn (b)}\right] \,, \label{TI}
\end{equation}
$$
T^{(II)}_{ik} = - \frac{1}{2}g_{ik}\left[ \D_{m}
\D_{l}\left(X_{(a)(b)}F^{mn (a)}F^{l (b)}_{\ n} \right) -
R_{lm}X_{(a)(b)}F^{mn (a)}F^{l (b)}_{\ n}\right] -
$$
$$
- F^{ln (a)} X_{(a)(b)}\left(R_{il}F^{(b)}_{kn} +
R_{kl}F^{(b)}_{in}\right) - \frac{1}{2} \D^m \D_m
\left(X_{(a)(b)}F^{(a)}_{in}F_{k}^{\ n (b)}\right) +
$$
\begin{equation}
+  \frac{1}{2}\D_l \left[ \D_i \left( X_{(a)(b)}F^{(a)}_{kn}F^{ln
(b)}\right) + \D_k \left(X_{(a)(b)}F^{(a)}_{in}F^{ln (b)}\right)
\right]  - R^{mn}X_{(a)(b)}F^{(a)}_{im}F^{(b)}_{kn} \,,
\label{TII}
\end{equation}
$$
T^{(III)}_{ik} = \frac{1}{4}g_{ik}
R^{mnls}X_{(a)(b)}F^{(a)}_{mn}F^{(b)}_{ls} - \frac{3}{4}
X_{(a)(b)}F^{ls (a)}\left(F_{i}^{\ n (b)} R_{knls} + F_{k}^{\ n
(b)}R_{inls}\right) -
$$
\begin{equation}
- \frac{1}{2}\D_{m} \D_{n} \left[X_{(a)(b)}\left(F_{i}^{\ n
(a)}F_{k}^{\ m (b)} + F_{k}^{\ n (a)}F_{i}^{\ m (b)}\right)
\right] \,, \label{TIII}
\end{equation}
$$
T^{(IV)}_{ik} = \left(R_{ik} - \frac{1}{2}g_{ik} R \right)
Y_{(a)(b)} (\D_m \Phi^{(a)})(\D^m \Phi^{(b)}) + R Y_{(a)(b)} (\D_i
\Phi^{(a)})(\D_k \Phi^{(b)}) +
$$
\begin{equation}
+ \left( g_{ik} \D^n \D_n - \D_i \D_k \right) \left[Y_{(a)(b)}
(\D_m \Phi^{(a)})(\D^m \Phi^{(b)}) \right] \,, \label{TIV}
\end{equation}
$$
T^{(V)}_{ik} =  Y_{(a)(b)} (\D_m \Phi^{(b)}) \left[ R^m_i (\D_k
\Phi^{(a)}) + R^m_k (\D_i \Phi^{(a)}) \right] - \frac{1}{2}R_{ik}
Y_{(a)(b)} (\D_m \Phi^{(a)})(\D^m \Phi^{(b)}) - {}
$$
$$
- \frac{1}{2} \D^m \left\{  \D_i \left[Y_{(a)(b)} (\D_m
\Phi^{(a)})(\D_k \Phi^{(b)}) \right] + \D_k \left[ Y_{(a)(b)}
(\D_m \Phi^{(a)})(\D_i \Phi^{(b)}) \right] - \right.
$$
\begin{equation}
\left.  - \D_m \left[ Y_{(a)(b)} (\D_i \Phi^{(a)})(\D_k
\Phi^{(b)}) \right] \right\} + \frac{1}{2}g_{ik} \D_m \D_n
\left[Y_{(a)(b)} \left(\D^m \Phi^{(a)}\right) \left(\D^n
\Phi^{(b)}\right) \right]
 \,. \label{TV}
\end{equation}

\section{Exact solution of the Wu-Yang type of the non-minimal EYMH
model}\label{ex}

\subsection{Ansatz about the fields structure}

Let the spacetime metric be of the spherically symmetric form
\begin{equation}\label{metrica}
ds^2=\sigma^2Ndt^2-\frac{dr^2}{N}-r^2 \left( d\theta^2 +
\sin^2\theta d\varphi^2 \right) \,.
\end{equation}
Here $\sigma$ and $N$ are functions depending on the radius $r$
only and satisfying the asymptotic  conditions
\begin{equation}\label{asymp}
\sigma\left(\infty\right)=1 \,,\quad N\left(\infty\right)=1 \,.
\end{equation}
We focus on the gauge field characterized by the special ansatz
(see, \cite{RebbiRossi}):
\[%
\mathbf{A}_{0}=\mathbf{A}_{r}=0 \,,
\]%
\begin{equation}\label{1}
\mathbf{A}_{\theta}= i \mathbf{t}_{(\varphi)},\quad
\mathbf{A}_{\varphi}=- i\nu \sin{\theta}\;\mathbf{t}_{(\theta)}\,.
\end{equation}
The parameter $\nu$ is a non-vanishing integer. The generators
${\bf t}_{(r)}$, ${\bf t}_{(\theta)}$ and ${\bf t}_{(\varphi)}$
are the position-dependent ones and are connected with the
standard generators of the SU(2) group as follows:
\[%
{\bf t}_{(r)}=\cos{\nu\varphi} \ \sin{\theta}\;{\bf
t}_{(1)}+\sin{\nu\varphi} \ \sin{\theta}\;{\bf
t}_{(2)}+\cos{\theta}\;{\bf t}_{(3)},
\]%
\begin{equation}%
{\bf t}_{(\theta)}=\partial_{\theta}{\bf t}_{(r)},\qquad {\bf
t}_{(\varphi)}=\frac {1}{\nu\sin{\theta}}\ \partial_{\varphi}{\bf
t}_{(r)} \,. \label{deS5}
\end{equation}%
The generators satisfy the relations
\begin{equation}%
\left[{\bf t}_{(r)},{\bf t}_{(\theta)}\right]=i\,{\bf
t}_{(\varphi)} \,,\quad \left[{\bf t}_{(\theta)} \,, {\bf
t}_{(\varphi)}\right]=i\,{\bf t}_{(r)} \,, \quad \left[{\bf
t}_{(\varphi)},{\bf t}_{(r)}\right]=i\,{\bf
t}_{(\theta)}\,.\label{deS6}
\end{equation}%
The field strength tensor
\begin{equation}\label{strength}
\mathbf{F}_{ik}=\partial_i\mathbf{A}_k-\partial_k\mathbf{A}_i+\left[\mathbf{A}_i
, \mathbf{A}_k\right]
\end{equation}
has only one non-vanishing component:
\begin{equation}\label{2}
{\bf F}_{\theta\varphi}= i \nu \sin\theta\,{\bf t}_{(r)}\,,
\end{equation}
which does not depend on the variable $r$. Our ansatz for the
Higgs field is the following: we consider ${\bf \Phi}$ as a
covariantly constant vector in the group space, which contains
only one component, i.e.,
\begin{equation}\label{das1}
{\bf \Phi} = \phi \,{\bf t}_{(r)} \,, \quad \D_k \Phi_{(a)} =
\partial_k \Phi_{(a)} + {\cal G} \varepsilon_{(a)(b)(c)} A^{(b)}_k
\Phi^{(c)} = 0 \,.
\end{equation}
This means that the Yang-Mills field strength tensor
$\mathbf{F}_{ik}$ and the Higgs field ${\bf \Phi}$ are parallel in
the group space \cite{Yasskin,Galtsov} to the vector ${\bf q} =
{\bf t}_{(r)}$. The Higgs equations are satisfied identically,
when the field $\phi$ is constant and coincides with $v$, the
background value of the scalar field given by
$\Phi^{(a)}\Phi_{(a)}=v^2$. Then the Yang-Mills equations
\Ref{HikR} are satisfied identically also, and we deal with a new
non-minimal Wu-Yang monopole solution, supplemented by the
background covariantly constant Higgs field.

Finally, the equations for the gravity field are extremely
simplified by the ansatz and its particular consequence
$X_{(a)(b)} F^{(b)}_{mn} = Q_1 F_{(a) mn}$. In fact the equations
for the gravity field for the model under discussion coincides
with the ones, described in \cite{2BZ06}, if we replace the
coupling parameters $q_1$, $q_2$, $q_3$ by the $Q_1q_1$, $Q_1q_2$,
$Q_1q_3$, respectively. Thus, we can use the exact solutions of
the Wu-Yang monopole type to the non-minimal Einstein-Yang-Mills
model, as an exact solutions for the generalized non-minimal EYMH
model with the covariantly constant Higgs field. The scalar field
in this case does not modify the pressure and the energy of the
system as whole, nevertheless, it play an important role, creating
the privilege direction in the group space. In particular, the
following solution for the non-minimal Wu-Yang monopole with
regular center
\begin{equation}\label{sN1}
\sigma(r)=1\,,\quad N=1+\frac{r^2\,(\kappa-4Mr)}{2\,(r^4+\kappa
q)} \,,
\end{equation}
is valid when $Q_1q_1=-q$, $Q_1q_2=4q$, $Q_1q_3=-6q$, and $q$ is
positive.

\section{Multi-metric representation of the material
tensors}\label{repr}

By analogy with decomposition of material tensor in electrodynamics
\cite{BZ05} the tensor $C^{ikmn}_{(a)(b)}$ can be represented as
\begin{equation}
C^{ikmn}_{(a)(b)} =
\frac{1}{2\hat{\mu}}\sum_{(\alpha)(\beta){(c)(d)}}
 G_{(\alpha)(\beta)(a)(b)}^{(c)(d)}\left(g^{im (\alpha)}_{(c)} \ g^{kn
(\beta)}_{(d)} - g^{in (\alpha)}_{(c)} \ g^{km (\beta)}_{(d)}
\right) \,, \label{supergeneral}
\end{equation}
with respect to color associated metrics $g^{im (\alpha)}_{(c)}$,
where $\hat{\mu}$ is some convenient factor \cite{BZ05}. Below we
apply the elaborated formalism to the case, when the non-minimal
Wu-Yang monopole accompanied by the constant scalar field forms
the gravitational background for the moving particles,
quasi-particles and waves.

\subsection{Associated metrics}\label{assoc}

In the spherically symmetric model under consideration the
material tensor $C^{ikmn}_{(a)(b)}$ reduces to
\begin{equation}
C^{ikmn}_{(a)(b)} = \delta^{(r)}_{(a)}\delta^{(r)}_{(b)}
C^{ikmn}_{({\rm long})} +
\left[\delta^{(\theta)}_{(a)}\delta^{(\theta)}_{(b)} +
\delta^{(\varphi)}_{(a)}\delta^{(\varphi)}_{(b)} \right]
C^{ikmn}_{({\rm trans})} \,, \label{as1}
\end{equation}
where $C^{ikmn}_{({\rm long})}$ and $C^{ikmn}_{({\rm trans})}$ are
defined by \Ref{lo1} and \Ref{tr1}, respectively. The
reconstruction of the tensor $C^{ikmn}_{({\rm long})}$ can be done
using two longitudinal color metrics $g^{ik (A)}_{({\rm long})}$
and $g^{ik (B)}_{({\rm long})}$ as follows (see \cite{BZ05} for
details)
$$
C^{ikmn}_{({\rm long})} = \frac{1}{2 \hat{\mu}_{({\rm long})}}
\left\{ \left[g^{im (A)}_{({\rm long})} g^{kn (A)}_{({\rm long})}
- g^{in (A)}_{({\rm long})} g^{km (A)}_{({\rm long})} \right] -
\right.
$$
\begin{equation}
\left. {} - \gamma_{({\rm long})} \left[\left(g^{im (A)}_{({\rm
long})}-g^{im (B)}_{({\rm long})}\right)\left(g^{kn (A)}_{({\rm
long})}-g^{kn (B)}_{({\rm long})}\right) - \left(g^{in (A)}_{({\rm
long})}-g^{in (B)}_{({\rm long})}\right)\left(g^{km (A)}_{({\rm
long})}-g^{km (B)}_{({\rm long})}\right) \right] \right\} \,,
\label{mainlong1}
\end{equation}
where
\begin{equation} g^{im (A)}_{({\rm long})} = U^{i} U^{m} +
\frac{\varepsilon_{\bot}}{\varepsilon_{||}} \Delta^{im} +
\frac{(\varepsilon_{\bot}-\varepsilon_{||})}{\varepsilon_{||}}
X^{i}_{\{r\}}X^{m}_{\{r\}} \,, \label{mainlong2}
\end{equation}
\begin{equation}
g^{im (B)}_{({\rm long})} = U^{i} U^{m} + \frac{\mu_{\bot}}{
\mu_{||}} \Delta^{im} + \frac{(\mu_{\bot}-\mu_{||})}{\mu_{||}}
X^{i}_{\{r\}}X^{m}_{\{r\}} \,. \label{mainlong3}
\end{equation}
Mention that $g^{ik (B)}_{({\rm long})}$ can be obtained from
$g^{ik (A)}_{({\rm long})}$ by a formal replacement of the symbol
$\varepsilon$ by the symbol $\mu$. We used the following notations
\begin{equation}\label{Bepsilon}
\varepsilon_{\bot} \equiv \varepsilon^{\theta}_{\theta} =
\varepsilon^{\varphi}_{\varphi}
 = 1 + 2 Q_1 {\cal R}^{\theta t}_{ \ \ \theta t}  = 1 + 2 Q_1 {\cal R}^{\theta r}_{ \ \ \theta
r} = (\mu^{-1})^{\theta}_{\theta} = (\mu^{-1})^{\varphi}_{\varphi}
\equiv \frac{1}{\mu_{\bot}} \,,
\end{equation}
\begin{equation}\label{Bmu}
\varepsilon^{r}_{r} \equiv \varepsilon_{||} = 1 + 2 Q_1 {\cal
R}^{r t}_{ \ \ r t}  \,, \quad (\mu^{-1})^{r}_{r} \equiv
\frac{1}{\mu_{||}} = 1 + 2 Q_1 {\cal R}^{\theta \varphi}_{ \ \
\theta \varphi} \,,
\end{equation}
\begin{equation}\label{Bg111}
\nu^{ \ k}_m = 0 \,, \quad \frac{1}{\hat{\mu}_{({\rm long})}} =
\varepsilon_{||} \,, \quad \frac{1}{\gamma_{({\rm long})}} = 1 -
\frac{\varepsilon_{||}\mu_{\bot}}{\varepsilon_{\bot}\mu_{||}} \,,
\end{equation}
\begin{equation}\label{ukh1}
U^k = \delta^k_t \frac{1}{\sqrt{N(r)}} \,, \quad X^k_{\{r\}} =
\delta^k_r \sqrt{N(r)} \,,
\end{equation}
\begin{equation}
{\cal R}^{r \theta}_{\ \ r \theta} = {\cal R}^{r \varphi}_{\ \ r
\varphi} = {\cal R}^{\theta t}_{\ \ \theta t} = {\cal R}^{\varphi
t}_{\ \ \varphi t} =  - \frac{r}{(r^4 + a^4)^3}\left[6q Mr^8 - 3a^4
r^7 - 24 q M a^4 r^4 +5 a^8 r^3 + 2q M a^8 \right] \,, \label{as4}
\end{equation}
\begin{equation}
{\cal R}^{r t}_{\ \ r t} =  \frac{r}{(r^4+a^4)^3}\left[12q M r^8 -
7 a^4 r^7 - 76 q M a^4 r^4 +17 a^8 r^3 + 8 q M a^8 \right] \,,
\label{as5}
\end{equation}
\begin{equation}
{\cal R}^{\theta \varphi}_{\ \ \theta \varphi} =
\frac{r^4}{(r^4+a^4)^3}\left[12q M r^5
 - 5 a^4 r^4 -20 q M a^4 r + 3 a^8 \right] \,. \label{as6}
\end{equation}
A new parameter $a$ with dimensionality of length is defined as
follows:  $a^4 = \kappa q$, where $\kappa = {8\pi\nu^2}/{{\cal
G}^2}$. To obtain the decomposition for $C^{ikmn}_{({\rm trans})}$
one can simply replace the mark $({\rm long})$ by $({\rm trans})$
and then replace $Q_1$ by one. Clearly, this simple model involves
the following non-trivial components of the tensor
$G_{(\alpha)(\beta)(a)(b)}^{(c)(d)}$, introduced by
\Ref{supergeneral}:
\begin{equation}
G_{(\alpha)(\beta)(\theta)(\theta)}^{(\theta)(\theta)} =
G_{(\alpha)(\beta)(\varphi)(\varphi)}^{(\varphi)(\varphi)} \neq
G_{(\alpha)(\beta)(r)(r)}^{(r)(r)} \,, \label{bigG}
\end{equation}
with $(\alpha), (\beta) = (A),(B)$, which satisfy the linear
relations
$$
G_{(A)(A)(a)(b)}^{(c)(d)}+ G_{(B)(A)(a)(b)}^{(c)(d)} = 1 \,, \quad
G_{(A)(B)(a)(b)}^{(c)(d)}+G_{(B)(B)(a)(b)}^{(c)(d)} = 0 \,,
$$
$$
G_{(A)(B)(r)(r)}^{(r)(r)} =
G_{(B)(A)(r)(r)}^{(r)(r)} = \gamma_{({\rm long})} \,,
$$
\begin{equation}
G_{(A)(B)(\theta)(\theta)}^{(\theta)(\theta)} =
G_{(B)(A)(\theta)(\theta)}^{(\theta)(\theta)} =
G_{(A)(B)(\varphi)(\varphi)}^{(\varphi)(\varphi)} =
G_{(B)(A)(\varphi)(\varphi)}^{(\varphi)(\varphi)}= \gamma_{({\rm
trans})} \,. \label{bigG3}
\end{equation}
Thus, in order to reconstruct the linear response tensor
$C^{ikmn}_{(a)(b)}$ for the model under discussion one needs four
associated metrics $g^{im (A)}_{({\rm long})}$, $g^{im (B)}_{({\rm
long})}$, $g^{im (A)}_{({\rm trans})}$ and $g^{im (B)}_{({\rm
trans})}$. When $Q_1=1$, i.e., there is no privilege direction in
the group space, the corresponding longitudinal and transversal
$A$ and $B$ metrics coincide.

\subsection{Color metrics}\label{color}

The associated metrics $g^{im (A)}_{({\rm long})}$, $g^{im
(B)}_{({\rm long})}$, $g^{im (A)}_{({\rm trans})}$ and $g^{im
(B)}_{({\rm trans})}$ are the color ones. To prove this statement
we have to check two facts. First, the equalities
\begin{equation}\label{Bas13}
g^{km (A)}_{({\rm long})} \ p_k p_m = 0 \,, \quad \text{or} \quad
g^{km (B)}_{({\rm long})} \ p_k p_m =0 \,,
\end{equation}
are satisfied in the WKB-approximation for A and B-waves,
respectively, when the Yang-Mills potential has a form
$A_k^{(r)}{\bf t}_{(r)}$. Second, the equalities
\begin{equation}\label{Bas14}
g^{km (A)}_{({\rm trans})} \ p_k p_m = 0 \,, \quad \text{or} \quad
g^{km (B)}_{({\rm trans})} \ p_k p_m =0 \,,
\end{equation}
hold, when the Yang-Mills potential is of the form
$A_k^{(\theta)}{\bf t}_{(\theta)}+A_k^{(\varphi)}{\bf
t}_{(\varphi)}$. Indeed, let us consider the Yang-Mills equations
in the leading order WKB-approximation:
\begin{equation}\label{WKB1}
C^{ikmn}_{(a)(b)} \ p_k p_m A^{(b)}_{n} = 0 \,,
\end{equation}
with the tensor $C^{ikmn}_{(a)(b)}$ given by \Ref{as1}, \Ref{lo1},
\Ref{tr1}. When the group index $(a)$ coincides with $(r)$, one
obtains
\begin{equation}\label{WKB2}
C^{ikmn}_{(r)(r)} \ p_k p_m A^{(r)}_{n} \equiv C^{ikmn}_{({\rm
long})} \ p_k p_m A^{(r)}_{n} = 0 \,.
\end{equation}
When $(a)=(\theta)$ or $(a)=(\varphi)$, the system \Ref{WKB1}
gives two equations for the transversal components of the
Yang-Mills potential $A^{(\theta)}_{n}$ and $A^{(\varphi)}_{n}$
\begin{equation}\label{WKB23}
C^{ikmn}_{({\rm trans})} \ p_k p_m A^{(a)}_{n} = 0 \,.
\end{equation}

In the WKB approximation the gauge potentials $A_k^{(a)}$ and the
field strength $F_{kl}^{(a)}$ can be extrapolated as follows
\begin{equation}
A_k^{(a)} \rightarrow {\cal A}_k^{(a)} e^{i \Psi} \,, \quad
F_{kl}^{(a)} \rightarrow i \left[ p_k {\cal A}_l^{(a)} - p_l {\cal
A}_k^{(a)}\right] e^{i \Psi} \,, \quad p_k = \nabla_k \Psi
\,.\label{AP1}
\end{equation}
Mention that the nonlinear terms in (\ref{Fmn}) give the values of
the next order in WKB approximation, thus, such a model of gauge
field is effectively Abelian. In the leading order approximation
the Yang-Mills equations reduce to
\begin{equation}
C^{ikmn}_{(a)(b)} \ p_k \ p_m \ {\cal A}_n^{(b)} = 0 \,.
\label{AP2}
\end{equation}
For the sake of simplicity and in this subsection only we omit the
marks (long) and (trans) in the associated metrics and in the
parameters $\mu$ and $\gamma$, since the results are similar for the
longitudinal and transversal cases. Substitution $C^{ikmn}_{(a)(b)}$
from (\ref{mainlong1}) with $g^{ik(A)}$ from (\ref{mainlong2}) and
$g^{ik(B)}$ from (\ref{mainlong3}) with $\hat{\mu}$ and $\gamma$
given by (\ref{Bg111}) yields
$$
 g^{im (A)} p_m  \left[g^{kn(A)}p_k {\cal A}_n^{(a)} \right] -
 g^{in (A)} {\cal A}_n^{(a)} \left[g^{km(A)}p_k p_m \right] =
$$
\begin{equation} \label{AP3}
= \frac{1}{\varepsilon^2_{||}\mu^2_{\bot}} \left(1-
\frac{\varepsilon_{||}\mu_{\bot}}{\varepsilon_{\bot}\mu_{||}}\right)
p_k p_m  {\cal A}_n^{(a)} \left[ \left(X^{i}_{(1)}X^{k}_{(2)} -
X^{i}_{(2)}X^{k}_{(1)} \right) \left(X^{m}_{(1)}X^{n}_{(2)} -
X^{m}_{(2)}X^{n}_{(1)} \right)\right] \,,
\end{equation}
where $X^i_{\{1\}}$, $X^i_{\{2\}}$ are unit spacelike vectors,
orthogonal to $U^i$, $X^i_{\{r\}}$, and each other. Projection of
this equations onto the velocity four-vector $U_i$ gives the scalar
ratio
\begin{equation} \label{AP4}
\left(U^m p_m \right) \left[g^{kn(A)} p_k {\cal A}_n^{(a)}\right] =
\left(U^n {\cal A}_n^{(a)}\right) \left[g^{km (A)} p_k p_m
\right]\,,
\end{equation}
which is satisfied if, for instance, we use the Landau gauge $U^n
{\cal A}_n^{(a)}=0$ and the condition of orthogonality of the wave
four-vector and amplitude four-vector in the first associated
metric, i.e., $g^{kn} p_k {\cal A}_n^{(a)}=0 $. Projections onto the
axes, given by $X^i_{\{1\}}$, $X^i_{\{2\}}$ and $X^i_{\{r\}}$ yield,
respectively,
$$
{\cal A}_{\{1\}}^{(a)} \left[ g^{km (A)} p_k p_m + p_{\{2\}}^2
\left(
 \frac{1}{\varepsilon_{||}\mu_{\bot}} - \frac{1}{\varepsilon_{\bot}\mu_{||}}\right) \right]-
 {\cal A}_{\{2\}}^{(a)} \left[ p_{\{1\}}p_{\{2\}} \left(
 \frac{1}{\varepsilon_{||}\mu_{\bot}} - \frac{1}{\varepsilon_{\bot}\mu_{||}}\right)
 \right]= 0 \,,
$$
$$
{\cal A}_{\{1\}}^{(a)} \left[ p_{\{1\}}p_{\{2\}} \left(
 \frac{1}{\varepsilon_{||}\mu_{\bot}} - \frac{1}{\varepsilon_{\bot}\mu_{||}}\right)
 \right] - {\cal A}_{\{2\}}^{(a)} \left[ g^{km (A)} p_k p_m + p_{\{1\}}^2 \left(
 \frac{1}{\varepsilon_{||}\mu_{\bot}} - \frac{1}{\varepsilon_{\bot}\mu_{||}}\right) \right]= 0 \,,
$$
\begin{equation} \label{AP5}
{\cal A}_{\{r\}}^{(a)} \left[g^{km (A)} p_k p_m \right]= 0 \,,
\end{equation}
where ${\cal A}_{\{1\}}^{(a)} \equiv X^k_{\{1\}}{\cal A}_{k}^{(a)}$,
$p_{\{1\}} \equiv X^k_{\{1\}}p_k$, etc. Taking into account the
relation
\begin{equation} \label{AP6}
\left[g^{km (B)} p_k p_m \right]= \left[g^{km (A)} p_k p_m \right] +
\left( p_{\{1\}}^2 + p_{\{2\}}^2\right)\left(
 \frac{1}{\varepsilon_{||}\mu_{\bot}} - \frac{1}{\varepsilon_{\bot}\mu_{||}}\right)
 \,,
\end{equation}
one can conclude that nontrivial solution of (\ref{AP5}) exists,
when
\begin{equation} \label{AP7}
\left[g^{km (A)} p_k p_m \right]\left[g^{lj (B)} p_l p_j \right]=
0 \,.
\end{equation}
Thus, the associated metrics $g^{km (A)}$ and $g^{km (B)}$ are the
color metrics, which indicate both longitudinal and transversal
ones.

\subsection{Color-acoustic metrics}\label{acoustic}

Tensor ${\cal C}^{ik}_{(a)(b)}$ \Ref{higgs111} can be associated
with two metrics $\tilde{g}^{ik}_{({\rm long})}$ and
$\tilde{g}^{ik}_{({\rm trans})}$:
\begin{equation}\label{ac1}
{\cal C}^{ik}_{(a)(b)} =  \left[
\delta^{(\theta)}_{(a)}\delta^{(\theta)}_{(b)} +
\delta^{(\varphi)}_{(a)}\delta^{(\varphi)}_{(b)}\right]
\tilde{g}^{ik}_{({\rm trans})} + \tilde{g}^{ik}_{({\rm long})}
\delta^{(r)}_{(a)}\delta^{(r)}_{(b)}\,.
\end{equation}
In the leading order of the WKB approximation, the propagation  of
the scalar particle with the color index $(a)= (r)$, described by
the potential $\Phi^{(r)} {\bf t}_{(r)}$, is equivalent to the
motion in the effective spacetime with the metric
$\tilde{g}^{ik}_{({\rm long})}$. As for the particle, described
the potential $\Phi^{(\theta)} {\bf t}_{(\theta)}$ or
$\Phi^{(\varphi)} {\bf t}_{(\varphi)}$, the corresponding
effective spacetime has a metric $\tilde{g}^{ik}_{({\rm trans})}$.
We obtain {\it color} birefringence for  scalar waves
(quasi-particles), since the velocity of the wave depends on its
color. Taking into account the formulas
\begin{equation}\label{ac3}
\Re^{t}_{t}= \Re^{r}_{r} = \frac{1}{2} (2q_4+q_5) N^{\prime
\prime}  + \frac{1}{r} (4q_4+q_5) N^{\prime} + \frac{2}{r^2} q_4
(N-1) \,,
\end{equation}
\begin{equation}\label{ac4}
\Re^{\theta}_{\theta} = \Re^{\varphi}_{\varphi}= q_4 N^{\prime
\prime} + \frac{1}{r} (4q_4+q_5) N^{\prime} + \frac{1}{r^2}
(2q_4+q_5) (N-1)\,,
\end{equation}
we can reconstruct the color-acoustic metric for arbitrary
parameters $Q_2$, $M$, $q_4$ and $q_5$. Nevertheless, to complete
the illustration of the possibilities of the effective metric
approach we restrict ourselves by the model with $M=0$. Then the
longitudinal color-acoustic metric is
\begin{equation}\label{ac5}
\tilde{g}^{ik}_{({\rm long})} = {\cal A}^2 \left\{ \left[
\frac{1}{N}\delta^{i}_{t}\delta^{k}_{t} - N
\delta^{i}_{r}\delta^{k}_{r} \right] - \frac{1}{r^2 {\cal B}^2}
\left[ \delta^{i}_{\theta}\delta^{k}_{\theta} +
\frac{1}{\sin^2\theta}\delta^{i}_{\varphi}\delta^{k}_{\varphi}
\right] \right\}\,,
\end{equation}
where the conformal factor ${\cal A}^2$ is
\begin{equation}\label{ac6}
{\cal A}^2 =  1 + \frac{Q_2}{2q (\xi+1)^3} \left[\xi^2 q_5 - 4\xi
(5q_4+3q_5) + 3(4q_4+q_5)\right] \,,
\end{equation}
and the angular factor ${\cal B}^2$ is given by
\begin{equation}\label{ac7}
{\cal B}^2 = \frac{2q (\xi+1)^3 + Q_2\left[\xi^2 q_5 - 4\xi
(5q_4+3q_5) + 3(4q_4+q_5) \right]}{2q (\xi+1)^3 + Q_2\left[- \xi^2
q_5 - 2\xi (10q_4-q_5) + 3(4q_4+q_5)\right] }\,.
\end{equation}
Here we introduced a new convenient dimensionless  variable $\xi
= r^4 / \kappa q$, which clearly takes non-negative values only.
Again, the transversal metric can be obtained from this formulas
after the formal replacement $Q_2 \to 1$.

\section{Discussion}

The formulas \Ref{mainlong2} and \Ref{mainlong3} present the
associated metrics for the non-minimal EYMH model under
discussion. As well, in the WKB approximation the Yang-Mills
equations are satisfied when \Ref{Bas13} or \Ref{Bas14} are valid.
Thus, these associated metrics can be interpreted as color ones.
This means, in particular, that the propagation of a test
Yang-Mills wave in the non-minimally active spacetime can be
described by two dispersion relations
\begin{equation}\label{om1}
\omega^2_{(A)} =  \frac{p^2_{\bot}}{\varepsilon_{||} \mu_{\bot} }
+ p^2_{||} \,, \quad \omega^2_{(B)} =
\frac{p^2_{\bot}}{\varepsilon_{\bot} \mu_{||} } + p^2_{||} \,,
\end{equation}
where
\begin{equation}\label{om2}
\omega = p_l U^l  \,, \quad p^2_{\bot} = - (p_{\theta} p^{\theta}
+ p_{\varphi} p^{\varphi}) \,, \quad p^2_{||} = - p_r p^r \,.
\end{equation}
When the wave propagates in the radial direction, i.e.,
$p_{\bot}=0$, or in the transverse one, i.e., $p_{||}=0$, the
corresponding phase velocities can be easily found as ${\cal
V}_{||} = \omega / p_{||}$, ${\cal V}_{\bot} = \omega / p_{\bot}$
For A-wave and B-wave they are, respectively,
\begin{equation}\label{om3}
{\cal V}^{(A)}_{||} = {\cal V}^{(B))}_{||} = 1 \,, \quad {\cal
V}^{(A)}_{\bot} =\frac{1}{\sqrt{\varepsilon_{||}\mu_{\bot}}} \,,
\quad {\cal V}^{(B)}_{\bot} =
\frac{1}{\sqrt{\varepsilon_{\bot}\mu_{||}}} \,,
\end{equation}
i.e., both the longitudinal ($Q_1\neq 1$) and transversal
($Q_1=1$) color waves, propagating radially, have the phase
velocity equal to the speed of light in the minimal vacuum.

As an illustration, consider now a specific particular case of the
exact solution \Ref{sN1} characterized by $M=0$. This solution
does not admit singularities of the horizon type, since $N>1$
everywhere. It is convenient to present the basic formulas for the
velocities of the waves propagating in the direction, orthogonal
to the radius, as follows
\begin{equation}\label{zom1}
\left({\cal V}^{(A)}_{\bot}(\xi)\right)^2
=\frac{1}{\varepsilon_{||}\mu_{\bot}} = \frac{\xi^3 +
(3+6Q_1)\xi^2 + (3-10Q_1)\xi +1}{\xi^3 + (3-14Q_1)\xi^2 +
(3+34Q_1)\xi +1}\,,
\end{equation}
\begin{equation}\label{zom2}
\left({\cal V}^{(B)}_{\bot}(\xi)\right)^2 =
\frac{1}{\varepsilon_{\bot}\mu_{||}}= \frac{\xi^3 + (3-10Q_1)\xi^2
+ (3+6Q_1)\xi +1}{\xi^3 + (3+6Q_1)\xi^2 + (3-10Q_1)\xi +1}\,.
\end{equation}
For these formulas one can obtain the transversal color wave
velocity when $Q_1=1$, as well as, for longitudinal one, say, when
$Q_1=2$. When do the phase velocities coincide with the speed of
light in minimal vacuum, i.e., when do they equal to one? The
following facts clarify this question.

\noindent (i) ${\cal V}^{(A)}_{\bot}(\infty) = {\cal
V}^{(B)}_{\bot}(\infty) = 1$ for arbitrary $Q_1$.

\noindent (ii) ${\cal V}^{(A)}_{\bot}(0) = {\cal
V}^{(B)}_{\bot}(0) = 1$ for arbitrary $Q_1$.

\noindent (iii) ${\cal V}^{(A)}_{\bot}(\frac{11}{5}) = 1$ for
arbitrary $Q_1$.

\noindent (iv) ${\cal V}^{(B)}_{\bot}(1) = 1$ for arbitrary $Q_1
\neq 2$.

\noindent (v) $\left({\cal V}^{(A)}_{\bot}(\xi)\right)^2>0$ for
any $\xi$, when $-0.241<Q_1<8/9$.

\noindent (vi) $\left({\cal V}^{(B)}_{\bot}(\xi)\right)^2>0$ for
any $\xi$, when $-54/25<Q_1<8/9$.

\noindent The behavior of the functions $\left({\cal
V}^{(A)}_{\bot}(\xi)\right)^2$ and $\left({\cal
V}^{(B)}_{\bot}(\xi)\right)^2$ depend on the value of the
parameter $Q_1$. Remember, that the value $Q_1=1$ relates to the
transversal color waves, and $Q_1 \neq 1$ relates to the
longitudinal one. In order to clarify the principal moments of
analysis we put further $Q_1=2$ for the longitudinal case. Let us
attract the attention to two details. The first one is connected
with the nulls of the functions under consideration, the second
relates to the infinite value. Both points relate to the
singularities of the color metrics. When the functions \Ref{zom1}
and \Ref{zom2} become negative, we deal with {\it inaccessibility
zones}.

\vspace{3mm} \noindent {\it Longitudinal A-wave}

\noindent The function $\left({\cal V}^{(A)}_{\bot}(\xi)\right)^2$
for $Q_1=2$ is negative for the ranges $ \sqrt{65}-8 < \xi < 1$ and
$ 3.283<\xi<21.731$. In this ranges of radial variable $\xi = r^4 /
\kappa q$ the propagation of the color waves of this type is
impossible. At $r=r_1 \equiv (\kappa q (\sqrt{65}-8))^{1/4}$ and
$r=r_2 \equiv (\kappa q)^{1/4}$ the phase velocity vanishes and the
wave stops. At $r=3.283$ and $r= 21.731$ the phase velocity is
infinite. Admissible ranges of $\xi$ for the wave propagation are,
therefore, $0<\xi<\sqrt{65}-8$, $1 < \xi < 3.283 $ and $21.731 < \xi
<\infty$. The first admissible range is of finite length, the
corresponding part of the space can be indicated as {\it first inner
admissible region} or {\it trapped region}. This trapped region is
arranged between two spheres of the zeroth radius and of the radius
$r_1$. On the left edge of this region ($r=0$) the phase velocity is
equal to one relating to the absence of classical gravity force
($N=1, N^{\prime}=0$). With increasing of radius the phase velocity
decreases monotonically and vanishes at the right edge of the
trapped region, thus, the transversally propagating wave stops at
this surface. The second ({\it inner}) admissible region can also be
indicated as trapped one, nevertheless, the phase velocity at the
left edge, $\xi=1$, is vanishing, whereas the right edge is
characterized by the infinite phase velocity, passing the value
equal to one at $\xi=11/5$. The third admissible range is infinite.
The phase velocity is infinite at the left edge of this region and
tends monotonically to one at infinity. The non-minimal radius $r=a$
indicates the boundary between the {\it first inaccessibility zone}
and second admissible region. The {\it second inaccessibility zone}
$3.283 <\xi < 21.731$ is characterized by the infinite phase
velocity barriers at the left and right edges.

\vspace{3mm} \noindent {\it Longitudinal B-wave}

\noindent When $Q_1=2$ the formula \Ref{zom2} reduces to
\begin{equation}\label{zom21}
\left({\cal V}^{(B)}_{\bot}(\xi)\right)^2 = \frac{\xi^2 -16 \xi -
1}{\xi^2 + 16 \xi - 1}\,,
\end{equation}
i.e., the denominator and numerator are presented by the
polynomials quadratic in $\xi$. Now this function is discontinuous
at $\xi = \sqrt{65}-8$, this point being the right edge of the
first admissible region, $0<\xi<\sqrt{65}-8$, which is the trapped
region of the second type. When the radius grows the phase
velocity increases from one to infinity, and a wave does not stop
at the boundary, but is, on the contrary, infinite. The range
$\sqrt{65}-8<\xi<\sqrt{65}+8$ relates to the inaccessible zone,
since $\left({\cal V}^{(B)}_{\bot}(\xi)\right)^2$ is negative. The
second admissible region is $\sqrt{65}+8<\xi<\infty$, the phase
velocity is equal to zero at $\xi = \sqrt{65}+8$ and tends to one
asymptotically at $\xi \to \infty$.

\vspace{3mm} \noindent {\it Transversal A-wave}

\noindent When $Q_1=1$, the curve $\left({\cal
V}^{(A)}_{\bot}(\xi)\right)^2$ has no discontinuity. The trapped
region of the first type is situated at $0<\xi<0.191$; the second
admissible region is arranged at $0.539<\xi<\infty$; curve passes
one at $\xi = 11/5$, then reaches the maximal value and tends to
one at $\xi \to \infty$. The inaccessibility zone is
$0.191<\xi<0.539$.

\vspace{3mm} \noindent {\it Transversal B-wave}

\noindent For this wave one has two trapped regions, two
inaccessibility zones and one infinite admissible region. The
first trapped region is characterized by the inequality
$0<\xi<0.191$, the phase velocity starts from one at the left edge
and reaches infinity at the right edge. The second one is at
$0.539<\xi<1.853$, the phase velocity starts from infinity at the
left edge and vanished at the right one. The first inaccessibility
zone is arranged at $0.191<\xi<0.539$, the phase velocity is
infinite at both edges. The second one is characterized by
$1.853<\xi<5.249$, the phase velocity being vanishing at both
edges. The infinite admissible region is $5.249<\xi<\infty$. The
phase velocity starts from zero at the left edge and tends
monotonically to one at $\xi \to \infty$.

\vspace{3mm} \noindent {\it Color-acoustic waves}

\noindent
In order to complete our illustration let us consider a
particular model with the following parameters:
\begin{equation}\label{ac8}
q_5 =  - \frac{q}{9} \,, \quad q_4 = \frac{q}{36} \,.
\end{equation}
Our choice relates to the idea that non-minimal radius $a=(\kappa
q)^{1/4}$, appeared in the equations for color gauge waves, is the
same one for the color-acoustic waves, i.e., there is no
additional non-minimal Constant of Nature (see, e.g.,
\cite{HDehnen4}). With \Ref{ac8} the functions ${\cal A}^2(\xi)$
and ${\cal B}^2(\xi)$ are, respectively,
\begin{equation}\label{ac9}
{\cal A}^2 =  \frac{18 (\xi+1)^3 + Q_2 \xi (7-\xi)}{18
(\xi+1)^3}\,,
\end{equation}
\begin{equation}\label{ac10}
{\cal B}^2 = \frac{18 (\xi+1)^3 + Q_2 \xi (7-\xi)}{18 (\xi+1)^3 -
Q_2 \xi (7-\xi)}\,,
\end{equation}
(see formulas \Ref{ac6} and \Ref{ac7}). The toy models with
$Q_2=2$ (longitudinal case) and $Q_2=1$ (transversal case) are
non-singular. Indeed, in both cases the cubic polynomials in the
numerators and denominators of \Ref{ac9} and \Ref{ac10} have no
real positive roots. This means that both functions, ${\cal
A}^2(\xi)$ and ${\cal B}^2(\xi)$ are continuous and positive for
$\xi>0$. In addition, both functions, start from one at $\xi=0$,
pass through one at $\xi=7$ and tend to one at $\xi \to \infty$.
As well, the function ${\cal B}^2(\xi)$ reaches the maximum at
$\xi = 8-\sqrt{57}$ and minimum at $\xi = 8+\sqrt{57}$. Thus, in
contrast to the case of color waves the color-acoustic waves in
this toy model have neither trapped regions, nor inaccessible
zones. The phase velocities of the corresponding color-acoustic
waves are non-monotonic functions of the radius, nevertheless,
they can reach neither zero value, nor infinity.

\vspace{3mm} \noindent {\it Remark}

\noindent In general case the propagation direction of the color
or color-acoustic wave is not pure longitudinal or pure
transversal with respect to the radial direction. In general case
the trajectory of the corresponding quasi-particle becomes very
sophisticated depending on the value of the impact parameter. This
requires a detailed numerical analysis, and we hope to discuss
that results in a separate paper.

\vspace{3mm} \noindent {\it Alternative description}

\noindent Particle dynamics can be alternatively described using
the equation of motion with effective force. Instead of equation
of null geodesics in the effective spacetime with metric
$g^{ik(\alpha)}_{(h)}$, where the upper index $(\alpha)$ indicates
the $A$ and $B$ cases in the classification of effective metrics,
the lower index $(h)$ is treated as $({\rm long})$ and $({\rm
trans})$, one can write the equation
\begin{equation} \label{AP31}
\frac{d^2 x^i}{d\tau^2} + \Gamma^i_{kl} \frac{dx^k}{d\tau}
\frac{dx^l}{d\tau} = F^{i (\alpha)}_{(h)} \,,
\end{equation}
where
\begin{equation} 
 F^{i (\alpha)}_{(h)} \equiv \Pi^{i(\alpha)}_{kl(h)} \frac{dx^k}{d\tau}
\frac{dx^l}{d\tau} \,, \quad \Pi^{i(\alpha)}_{kl(h)} \equiv
\Gamma^i_{kl} - \Gamma^{i(\alpha)}_{kl(h)} \,.
\end{equation}
$\Gamma^i_{kl}$ and $\Gamma^{i(\alpha)}_{kl(h)}$ are the
Christoffel symbols for the real and effective spacetimes,
respectively. The quantity $\Pi^{i(\alpha)}_{kl(h)}$, the
difference of the Christoffel symbols, symmetric with respect to
indices $k$ and $l$, is known to be a tensor, thus, the quantity
$F^{i (\alpha)}_{(h)}$ is a vector. Since we consider the
interaction of Yang-Mills and Higgs field with spacetime
curvature, we can indicate this force as a tidal one. The tidal
force $F^{i (\alpha)}_{(h)}$ is quadratic in the particle
four-velocity and is predetermined by the structure of the tensor
$\Pi^{i(\alpha)}_{kl(h)}$.

For the Wu-Yang model model the tensor $\Pi^{i(A)}_{kl({\rm
long})}$ has the following non-vanishing components:
$$
\Pi^{r(A)}_{\theta \theta ({\rm long})} = \Pi^{r(A)}_{\varphi
\varphi ({\rm long})} / \sin^2\theta = \frac{1}{2} N(r)
\frac{d}{dr}\left[r^2
\left(\frac{\varepsilon_{||}}{\varepsilon_{\bot}} - 1 \right)
\right] \,,
$$
\begin{equation} \label{AP33}
\Pi^{\theta (A)}_{\theta r ({\rm long})} =
\Pi^{\varphi(A)}_{\varphi r ({\rm long})} = \frac{1}{2}
\frac{d}{dr}\left[\ln{
\left(\frac{\varepsilon_{||}}{\varepsilon_{\bot}} \right)} \right]
\,.
\end{equation}
The non-vanishing components of the tensor $\Pi^{i(B)}_{kl ({\rm
long})}$ can be obtained from \Ref{AP33} by the formal replacement
of the symbol $\varepsilon$ by the symbol $\mu$. As for
$\Pi^{i(A)}_{kl ({\rm trans})}$ and $\Pi^{i(B)}_{kl ({\rm
trans})}$, they can be obtained from the corresponding
longitudinal components if we put $Q_1 \to 1$. When the particle
moves radially, i.e., $\frac{d\theta}{d\tau} = \frac{d
\varphi}{d\tau} = 0$, the tidal force vanishes. This agrees with
the result obtained above that the radially propagating color
waves have the phase velocity equal to the speed of light in the
standard (minimal) vacuum.

\appendix

\begin{acknowledgments}
The authors are grateful to Prof. W. Zimdahl for the fruitful
discussion. This work was supported by the Deutsche
Forschungsgemeinschaft through project No. 436RUS113/487/0-5.
\end{acknowledgments}

\end{document}